\documentclass[sigconf,final]{acmart}
\RequirePackage[T1]{fontenc}
\usepackage{xr}
\usepackage{nameref}
\usepackage{tcolorbox}
\usepackage{color,soul}
\usepackage{flushend}
\usepackage{mdframed}
\usepackage{multirow,pbox,slashbox,tabularx,booktabs}
\usepackage[export]{adjustbox}
\usepackage{amsmath,amsfonts,amssymb,amsthm,paralist}
\usepackage{algorithmicx}
\usepackage[normalem]{ulem}
\usepackage{enumitem}
\usepackage{graphicx}
\usepackage{array}
\usepackage{xcolor}
\usepackage{hyperref}
\usepackage{pdfpages}

\copyrightyear{2020}
\acmYear{2020}
\setcopyright{acmcopyright}\acmConference[JCDL '20]{Proceedings of the ACM/IEEE Joint Conference on Digital Libraries in 2020}{August 1--5, 2020}{Virtual Event, China}
\acmBooktitle{Proceedings of the ACM/IEEE Joint Conference on Digital Libraries in 2020 (JCDL '20), August 1--5, 2020, Virtual Event, China}
\acmPrice{15.00}
\acmDOI{10.1145/3383583.3398512}
\acmISBN{978-1-4503-7585-6/20/06}

\usepackage[textsize=scriptsize]{todonotes}
\RequirePackage{colortbl}

\graphicspath{{../}}
\author{Jagriti Jalal}
\affiliation{%
  \institution{IIT Kharagpur, India}
  }
\email{jagritijalal@iitkgp.ac.in}

\author{Mayank Singh}
\affiliation{%
  \institution{IIT Gandhinagar, India}
  }
\email{singh.mayank@iitgn.ac.in}

\author{Arindam Pal}
\affiliation{%
  \institution{Data61, CSIRO \& Cyber Security CRC
  \\Sydney, New South Wales, Australia}
}
\email{arindamp@gmail.com}

\author{Lipika Dey}
\affiliation{%
  \institution{TCS Innovation Labs, India}
  }
\email{lipika.dey@tcs.com}

\author{Animesh Mukherjee}
\affiliation{%
  \institution{IIT Kharagpur, India}
  }
\email{animeshm@cse.iitkgp.ac.in}


\begin{document}

\title{Identification, Tracking and Impact:\\ Understanding the trade secret of catchphrases}

\begin{abstract}
Understanding the topical evolution in industrial innovation is a challenging problem.  With the advancement in the digital repositories in the form of patent documents, it is becoming increasingly more feasible to understand the innovation secrets -- `catchphrases' -- of organizations. However, searching and understanding this enormous textual information is a natural bottleneck. In this paper, we propose an unsupervised method for the extraction of catchphrases from the abstracts of patents granted by the U.S. Patent and Trademark Office over the years. Our proposed system achieves substantial improvement, both in terms of precision and recall, against state-of-the-art techniques.  As a second objective, we conduct an extensive empirical study to understand the temporal evolution of the catchphrases across various organizations. We also show how the overall innovation evolution in the form of introduction of newer catchphrases in an organization's patents correlates with the future citations received by the patents filed by that organization. Our code and data sets will be placed in the public domain.
\end{abstract}

\maketitle
\section{Introduction}
\label{sec:intro}
As software and other products are becoming more complex, the number and size of patent documents are increasing gradually. Automated patent document processing systems are essential to extract information and gain insights from this ever-increasing collection of patent databases. Catchphrases provide a concise representation of the content of a document. A catchphrase is a well-known word or phrase encapsulating the particular concept or subject of a document.  They contain all the important legal and technical aspects, instead of just summarizing the document. They have numerous applications such as document categorization,  clustering, summarization, indexing, topic search, quantifying semantic similarity with other documents, and conceptualizing particular knowledge domain of the document~\cite{gopavarapu2016increasing, jones1999topic}. However, since only a small minority of documents have author-assigned catchphrases, and manual assignment of catchphrases to existing documents is time-consuming, the automation of the catchphrase extraction process is highly desirable. In the current study, catchphrases represent innovation topics. Figure~\ref{fig:example_catchphrases} presents example catchphrases from two different patent abstracts.

In this paper, we propose an unsupervised method for the extraction of catchphrases from the abstracts of patents granted by the U.S. Patent and Trademark Office over the years. The key contributions of this paper are as follows.
\begin{itemize}
\item We propose an unsupervised technique for catchphrase identification and ranking in patent documents. 
\item We conduct robust evaluations and comparison against several state-of-the-art baselines. 
\item As a secondary objective, we study the evolution of catchphrases present in the patents filed by various organizations over time. 
\item We bring forth some of the unique temporal characteristics of these catchphrases and show how these are correlated to the overall future citation count of the patents filed by an organization. 
\item The catchphrase evolution study further unfolds that companies get polarized based on whether the patent documents keep re-using the same catchphrases over time or they introduce newer catchphrases as time progresses.
\end{itemize}

\begin{figure}[!tbh]
    \centering
    \begin{tcolorbox}
\textbf{ID:} US06681004\\
\textbf{Abstract}: The telephone memory aid provides a database to a primary party for storing and retrieving \hl{personal information} about a secondary party, including \hl{summary} information related to \hl{communication exchanges} between the primary and secondary parties. The summary information includes, for example, the date and time of prior telephone calls and the topics discussed. This secondary party information, including the summaries of prior telephone calls, is available for review by the primary party during future phone calls with the secondary party. The telephone memory aid also facilitates entry of information into the \hl{database} through \hl{speech recognition} algorithms and through \hl{question} and answer \hl{sessions} with the primary and secondary parties.
\vspace{0.5cm}\\
\textbf{ID:} US06680003\\
\textbf{Abstract}: The present invention concerns \hl{chiral doping agents} allowing a modification to be induced in the spiral pitch of a \hl{cholesteric liquid crystal}, said doping agents including a \hl{biactivated chiral unit} at least one of whose functions allows a chemical link to be established with an isomerisable group, for example by \hl{radiation}, said group possibly having a polymerisable or co-polymerisable end chain. These new chiral \hl{doping agents} find application in particular in a \hl{color display}.
\end{tcolorbox}
    \caption{Example abstracts from USPTO patents US06681004 and US06680003. The highlighted set of words are identified as catchphrases from IPC (described in Section~\ref{sec:exp}).}
    \label{fig:example_catchphrases}
\end{figure}

\section{Related Work}
A variety of techniques have been applied for automated keyword extraction like locating important phrases by analyzing markups like capitalization, section headings and emphasized texts~\cite{krulwich1997infofinder}; building phrase dictionary by parts-of-speech (POS) tagging of word sequences~\cite{larkey1999patent}; thesaurus-based keyphrase indexing~\cite{witten2006thesaurus}; domain-specific keyphrase extraction~\cite{frank1999domain,nguyen2009ontology} and several other supervised methods such as KEA~\cite{witten2005kea}, MAUI~\cite{medelyan2009human},  back-of-the-book indexing using catchphrase extraction~\cite{csomai2008linguistically}, MAUI with text denoising~\cite{shams2012investigating}, CSSeer~\cite{chen2013csseer} etc. In recent years artificial neural networks (ANNs) are being used to build predictive models to rank words in a document~\cite{boger2001automatic} and then select keywords based on these ranks.\par
\noindent It has been widely recognized that the innovative capability of a firm is a critical determinant of its performance and competitive edge~\cite{bettis1995new,helfat2003dynamic,greenhalgh2005running}. Since patents are a direct outcome of the inventive process and are broken down by technical fields, they are considered indicators of not only the rate of the innovative activities of a firm but also its direction~\cite{bloom2002patents,archibugi1996measuring,artz2010longitudinal}. Many previous studies have examined the relationships between the patenting activities of a company and its market value~\cite{oh2012evaluating,hall2005market}.~\citet{bornmann2008citation} precisely reviews the citing behavior of scientists and shows the role of citations as a reliable measure of impact. ~\citet{cheng2009profitability} shows that some indicators of patent quality are statistically significant to return on assets. ~\citet{lee2012stochastic} assesses future technological impacts by employing the future citation count as a proxy while~\citet{lee2018early} employs various patent indicators such as novelty and scope, as features of an ANN for early identification of emerging technologies.

\section{Datasets and Preprocessing}
\label{sec:data}
The current study requires a rich time-stamped dataset. We, therefore, leverage two independent data sources.
These are:
\begin{enumerate}
    \item \textbf{The patent dataset}: We compile the first dataset by crawling the full-text patent articles, available at the United States Patent and Trademark Office (USPTO\footnote{\url{https://bulkdata.uspto.gov/}}). It comprises patents granted weekly (Tuesdays) from January 1, 2003, to May 18, 2018 (excluding images/drawings). The patents are available as XML encoded files with English as the primary language. Out of all the curated documents, in this study, we only consider those patents for which the abstract information is present (see Table~\ref{tab:dataset} for statistics).
    \item \textbf{The newsgroup corpus}: We also use another data source, the \emph{20 Newsgroups Dataset}\footnote{\url{https://archive.ics.uci.edu/ml/datasets/Twenty+Newsgroups}} donated by T. Mitchell in 1999. It includes one thousand Usenet articles each from 20 newsgroups like 'alt.atheism', 'comp.graphics', 'talk.politics.guns', etc. Approximately 4\% of the articles are crossposted.\par
    \noindent This serves as a non-patent corpus to estimate the importance of a word specifically in the domain of the patents concerning a non-patent domain (see Table~\ref{tab:dataset} for statistics). 
    \end{enumerate}

\begin{table}[ht]
\centering
  \begin{tabular}{llc} \toprule
   \parbox[t]{2mm}{\multirow{3}{*}{\rotatebox[origin=c]{90}{\textbf{Patent}}}}&Year range& 2003--2018 \\ 
  &Number of patents &3,915,639 \\
  &Number of patents with abstract & 3,486,866 \\\hline
  \parbox[t]{2mm}{\multirow{4}{*}{\rotatebox[origin=c]{90}{\textbf{Newsgroup}}}}&Year range& 1993--2017\\ 
  &Number of articles &19,997\\ 
  & Number of words & --\\
  &Language& English\\
  \bottomrule \hline
  \end{tabular}
 \caption{\label{tab:dataset}General statistics about the patent dataset and the newsgroup corpus. A large fraction (89\%) of patents have abstract information.}
\end{table}

\noindent\textbf{Pre-processing}: For both of the above, we performed several pre-processing tasks such as a sentence to lowercase conversion, removal of special characters except apostrophe and periods, lemmatization, and multiple white-spaces removals.

\section{Catchphrase extraction}
\label{sec:method}
Catchphrase extraction is a challenging problem mainly due to the diversity and unavailability of large-text annotated datasets. We, therefore, present an unsupervised method for catchphrase extraction. We propose a two-stage extraction strategy that identifies relevant candidate catchphrases in a given patent article. In the first stage, we select the candidate catchphrases. This is followed by candidate catchphrase ranking in the second stage. Next, we describe the two stages in detail.

\subsection{Stage-1: Candidate selection}
In the first stage, we select candidate catchphrases from each patent's abstract. Empirically, we observe that all catchphrases are n-gram noun phrases, for example, unigrams (e.g. \textit{communication, dielectrometry, etc.}), bigrams (e.g. \textit{consecutive bit, voice synthesizer, etc.}), trigrams (e.g. \textit{integrated circuit device, hydrogen chloride gas, etc.}) or quadrigrams (e.g. \textit{commercially available synthesis tool, electric signal processing board, etc.}). We, therefore, perform part-of-speech-tagging (POS) of each abstract text to identify noun phrases. Currently, we leverage python's state-of-the-art NLP library SpaCy\footnote{\url{https://spacy.io/usage/linguistic-features}}. 
Note that, we experimented with two text processing approaches before noun phrase identification: (i) with stopwords (WS), and (ii) without stopwords (WOS). WS represents that no stopwords were removed from the abstracts, whereas, WOS represents that all stopwords in the abstract text were removed beforehand. Abstracts with stopwords (WS) led to better quality extraction results due to the existence of stop-words in noun phrases. We discuss the results in detail in Section~\ref{sec:exp}. Table~\ref{tab:stage1} presents statistics of extracted candidate phrases from the dataset. 

\begin{table}[!tbh]
  \begin{center}
    \begin{tabular}{l|c}
      \textbf{Word n-grams} & \textbf{Count}\\
      \hline
      Unigrams & 208,105\\
      Bigrams & 2,616,762\\
      Trigrams & 4,432,251\\
      Quadrigrams & 2,138,696\\ \hline
      \textbf{Total}&\textbf{9,395,814}\\
    \end{tabular}
    \caption{\label{tab:stage1}Count of n-gram noun phrases generated from patent dataset.} 
  \end{center}
\end{table}

\subsection{Stage-2: Candidate ranking}
\label{sec:scoring}
Candidate phrases obtained in the first stage are ranked in this stage. The ranking algorithm is based upon two empirical findings: (i) \textit{how well the phrase describes the document's topic}, and (ii) \textit{how specific is the phrase to the patent literature}. Our proposed method unifies both of these findings by combining a frequency-based measure with an information-theoretic measure. Given a patent document $d$ and a set of candidate phrases $c_d$ obtained in the previous stage, we compute the phrase score $PS(c,d)$ for each phrase $c \subseteq c_d$.

\begin{equation}\label{eqn1}
PS(c,d) = \sum_{i=1}^{\mid c \mid}\{log(score(t_i))\}.KLI(c, d) 
\end{equation}

where, $t_i$ denotes the $i^\textrm{th}$ term in an n-gram candidate phrase $c$, score($t_i$) denotes the score of the $i^\textrm{th}$ term by estimating the importance of the term specifically in the patent domain relative to a non-patent domain and $KLI(c,d)$ represents the Kullback-Leibler divergence informativeness specifying how well a candidate phrase $c$ represents a document $d$. The term $score(t)$ in the above equation is computed as
\begin{equation}
\label{eq:2}
   score(t) = \frac{Importance(t, C_p)}{Importance(t, C_{n})+1}
\end{equation}

Again, here, $C_p$ and $C_n$ represents the patent collection and non-patent (in our case, the newsgroup) collection. The importance$(t, C)$ of a term $t$ in a given collection $C\in \{C_p,C_n\}$ is measured in terms of the collection frequency $CF$ and the document frequency $DF$. $CF(t,C)$ represents how many times the term $t$ appeared in the entire collection $C$. $DF(t,C)$ represents the count of documents where the term $t$ appeared. It is computed as

\begin{equation}
       Importance(t, C) = \frac{CF(t, C)}{DF(t, C)+1}
\end{equation}
 
$KLI(c, d)$ denotes an information theoretic measure to compute  how informative the phrase is in the given document $d$. It is computed as: 
\begin{equation}
\label{eq:4}
       KLI(c, d) = \frac{TF(c,d)}{\mid d \mid}.\log{\frac{\frac{TF(c,d)}{\mid d \mid}}{\frac{CF(c)}{n}}}
   \end{equation}

where, $TF(c,d)$ represents how many times $c$ appeared in document $d$. $CF(c)$ denotes how many times $c$ appeared in the entire patent collection $C_p$. $|d|$ and $|n|$ represents total number of n-grams in document $d$ and $C_p$ respectively. 

The above scoring method results in a ranking of candidate phrases. We select top-ranked candidates such as top-5, top-10, top-20, etc., and evaluate our unsupervised method in the next section.

\section{Experiments}
\label{sec:exp}
In this section, we describe the experimental settings, baselines and the evaluation metrics. We construct a collection of possible catchphrases from the International Patent Classification (IPC) list. This list is maintained by the \textit{World Intellectual Property Organization} (WIPO)\footnote{\url{http://www.wipo.int/classifications/ipc/ipcpub/}}. The IPC provides a hierarchical system of language independent symbols for the classification of patents and utility models according to the different areas of technology to which they pertain. The hierarchy comprises eight high-level categories: 
\begin{enumerate}
    \item \textbf{Cat-1}:  Human necessities
    \item \textbf{Cat-2}:  Performing operations; Transporting
    \item \textbf{Cat-3}:  Chemistry; Metallurgy
    \item \textbf{Cat-4}:  Textiles; Paper
    \item \textbf{Cat-5}:  Fixed constructions
    \item \textbf{Cat-6}:  Mechanical engineering; Lighting; Heating; Weapons; Blasting
    \item \textbf{Cat-7}:  Physics 
    \item \textbf{Cat-8}:  Electricity
\end{enumerate}

In each of these high-level categories, several sub-categories exist. An n-gram phrase represents each category. We term these phrases as \textit{ground truth catchphrases} (GTC). Overall, we obtained 22,855 GTC such as ''actuators'', ''cleaning fabrics'', ''feedback arrangements in control systems'', etc. We use GTC to evaluate our proposed catchphrase extraction method. Table~\ref{tab:GT_example} presents examples of GTC for each high-level category. Next, we present three state-of-the-art baselines. 

\begin{table*}[!tbh]
    \centering
    \begin{tabular}{c|c|c|c|c}\toprule
        Category &Unigrams &Bigrams &Trigrams &Quadgrams \\\hline
        Cat-1 &rhinoscopes &dental surgery &table service equipment &foodstuffs containing gelling agents \\
        Cat-2 &thwarts &rivet hearths &making plough shares &making plastics bushes bearings \\
        Cat-3 &riboflavin &septic tanks &acetone carboxylic acid &chromising of metallic material surfaces \\
        Cat-4 &carding &carbon filaments &opening fiber bales &drying wet webs in paper-making \\
        Cat-5 &collieries &suspension bridges &setting anchoring bolts &freezing for sinking mine shafts \\
        Cat-6 &thermal &diesel engines &portable accumulator lamps &treating internal-combustion engine exhaust \\
        Cat-7 &ozotypy &investigating abrasion &measuring electric supply &incineration of solid radioactive waste \\
        Cat-8 &rheostats &electric accumulator &thermo magnetic devices &electric amplifiers for amplifying pulse \\\bottomrule
    \end{tabular}
    \caption{\label{tab:GT_example}Examples of ground truth catchphrases for each high-level category available in the International Patent Classification (IPC) list.}
\end{table*}

\subsection{Baselines}
\label{sec:baseline}
\begin{enumerate}
\item \textbf{Keyphrase extraction algorithm (KEA)}: KEA~\cite{witten2005kea} is a supervised machine learning toolkit that extracts keyphrases and ranks them. The original algorithm was trained on scientific documents and uses a trained  Na\"ive Bayes model. We trained KEA for patent documents leveraging a similar training procedure. 
\item \textbf{Legal}:~\citet{mandal2017automatic} also follow an unsupervised approach for identification of catchphrases from legal court cases. The scoring is done as:
    \begin{equation}\label{eqn5}
        PS(c, C_p, C_{np}) = \log[\sum_{i=1}^{\mid c \mid}Score(t_i, C_p, C_{np})].KLI(c, d) 
    \end{equation}
        where $score(t_i, C_p, C_np)$ and $KLI(c, d)$ can be calculated using equations \ref{eq:2} and \ref{eq:4} respectively. Note the change in the formula in equation~\ref{eqn5} compared to equation~\ref{eqn1}. This modification as we shall see almost doubles our performance. 
\item \textbf{KLIP}:~\citet{tomokiyo2003language, verberne2016evaluation} proposed a Kullback-Leibler (KL) divergence based phrase assignment score which is a linear combination of two different scores:
    \begin{enumerate}
        \item \textit{KL informativeness} (\textit{KLI}): KLI measures how well a candidate phrase represents a document. It is computed using equation~\ref{eq:4}.
        \item \textit{KL phraseness} (\textit{KLP}): KLP score is computed specifically for multi-word phrases. It compensates for low frequency of multi-word phrases by assigning higher weights to longer phrases:
\begin{equation}
    KLP(c,d) = \frac{TF(c,d)}{\mid d \mid}.\log{\frac{\frac{TF(c,d)}{\mid d \mid}}{\prod_{i=1}^{\mid c \mid} \frac{freq(t_i,d)}{\mid d \mid}}}
\end{equation}
    \end{enumerate}
where, $t_i$ is the $i^\textrm{th}$ term of the phrase $c$, and $freq(t_i,d)$ is the frequency of the term $t_i$ in document $d$.
        
\item \textbf{BM25}: BM25~\cite{robertson2009probabilistic} is a well-known measure for scoring documents with respect to a given query. We use this function for assigning score to an extracted candidate phrase $c$ in a given document $d$. The scoring function is:
\begin{equation}
    score(c, d) = IDF(c).\frac{TF(c,d).(k_1+1)}{TF(c,d)+k_1.\left(1+b+b.\frac{\mid d \mid}{avgdl}\right)}
\end{equation}
where $TF(c,d)$ is the term frequency of phrase $c$ in the document $d$. $k_1$ and $b$ are free parameters. We choose $k_1$ $\in$ [1.2, 2.0] and $b$ = 0.75\footnote{We select these values as per previous literature~\cite{robertson2009probabilistic}.}. $IDF(c)$ is the inverse document frequency of the candidate phrase $c$, calculated as
\begin{equation}
    IDF(c) = \log {\frac{n-DF(c)+0.5}{DF(c)+0.5}}
\end{equation}
where $DF(c)$ is the document frequency of the phrase $c$ in the collection.
\end{enumerate}
Note that KEA is a supervised machine learning model whereas Legal, KLIP and BM25 are unsupervised methods.  

\subsection{Evaluation measures} We evaluate our proposed method against the three baselines. We use two standard evaluation measures: (i) \textbf{Macro precision}, and (ii) \textbf{Macro recall}. These metrics are computed by macro-averaging the precision/recall values computed for every patent.
\begin{equation}
\text{Macro precision} = \frac{\sum_{i=1}^{\mid T \mid}  precision_i}{\mid T \mid}
\end{equation}
\begin{equation}
\text{Macro recall} = \frac{\sum_{i=1}^{\mid T \mid} recall_i}{\mid T \mid}
\end{equation}
where $precision_i$ and $recall_i$ are the precision and recall values computed for $i^\textrm{th}$ patent in our test dataset $T$. The precision and recall values for the $i^\textrm{th}$ patent are computed as follows 

\begin{equation}
precision_i = \frac{DCP_i}{DC_i}
\end{equation}

\begin{equation}
recall_i = \frac{DCP_i}{CPG_i}
\end{equation}

where $DC_i$, $DCP_i$, and $CPG_i$ represents the number of catchphrases in the $i^\textrm{th}$ patent that are detected, detected and present in GTC, and present in GTC respectively.

As KEA requires training, we partition our dataset into two classes: (i) train and (ii) test. Train split consist of 2,055,588 (65\%) patent documents. Test split consist of 1,106,883 (35\%) patent documents. For a fair comparison, we evaluate our proposed method against baselines (described in Section~\ref{sec:baseline}) using only the test split.

\subsection{Results and discussion}
\label{sec:res}
Table~\ref{tab:results} compares our proposed catchphrase extraction approach against state-of-the-art baselines. We outperform all baselines by a substantially high margin. The second best system in terms of precision is KEA, whereas the second-best system in terms of recall is a mix between KLIP and KEA. The baseline Legal performed worst among all the baselines, which is possible because of the fact that the authors take a logarithm of the sum of all the scores rather than the sum of the logarithms of the scores. The former measure undermines the contribution of the scores from each term and is therefore ineffective and is rather unintuitive. 

\begin{table*}[t]
    \centering
    \begin{tabular}{c| *{4}{c|} c || *{4}{c|} c }
    \multirow{2}{*}{\textbf{z}} & \multicolumn{5}{c||}{PRECISION} & \multicolumn{5}{c}{RECALL} \\
       & \textbf{Our Model} & \textbf{KEA}& \textbf{Legal} & \textbf{BM25} & \textbf{KLIP}  &  \textbf{Our Model} & \textbf{KEA}& \textbf{Legal} & \textbf{BM25} & \textbf{KLIP} \\
        \hline
      10 & \cellcolor{green!20}\textbf{0.253} & \cellcolor{yellow!20}0.192 & \cellcolor{red!20}0.075 & 0.080 & 0.120 & \cellcolor{green!20}\textbf{0.773} & \cellcolor{yellow!20}0.557 & \cellcolor{red!20}0.255 & 0.265 & 0.386 \\
      15 & \cellcolor{green!20}\textbf{0.231} & \cellcolor{yellow!20}0.148 & \cellcolor{red!20}0.128 & 0.131 & 0.146 & \cellcolor{green!20}\textbf{0.910} & 0.566 & \cellcolor{red!20}0.559  & 0.567 & \cellcolor{yellow!20}0.623 \\
      20 & \cellcolor{green!20}\textbf{0.217} & \cellcolor{yellow!20}0.133 & \cellcolor{red!20}0.156 & 0.156 & 0.164 & \cellcolor{green!20}\textbf{0.945} & 0.568 & \cellcolor{red!20}0.750 & 0.749 & \cellcolor{yellow!20}0.772 \\
      15\% & \cellcolor{green!20}\textbf{0.260} & \cellcolor{yellow!20}0.200 & \cellcolor{red!20}0.056 & 0.060 & 0.108 & \cellcolor{green!20}\textbf{0.750} & \cellcolor{yellow!20}0.555 & \cellcolor{red!20}0.172 & 0.185 & 0.323 \\
      20\% & \cellcolor{green!20}\textbf{0.240} & \cellcolor{yellow!20}0.156 & \cellcolor{red!20}0.109 & 0.111 & 0.132 & \cellcolor{green!20}\textbf{0.886} & \cellcolor{yellow!20}0.563 & \cellcolor{red!20}0.448 & 0.457 & 0.528 \\\hline
    \end{tabular}
    \caption{\label{tab:results}Comparison of our proposed method against the baselines: Precision and recall values at different top-ranks ($z\in{10,15,20,15\%,20\%}$) of extracted catchphrases.}
\end{table*}

\section{Temporal study}
\label{sec:temp_study}
In this section, we intend to show the usability of catchphrase extraction. We claim that catchphrase evolution presents a fair understanding of the changing innovation trends of companies. We conduct several interesting temporal studies to understand the emergence of new research topics in the industry. In this study, we  select top-10 companies from three industrial segments: (1) Software\footnote{\url{https://en.wikipedia.org/wiki/List_of_the_largest_software_companies}}, (2) Hardware\footnote{\url{https://www.investopedia.com/articles/investing/012716/worlds-top-10-hardware-companies-aaplibm.asp}},  and (3) Mobile Phones\footnote{\url{https://www.researchsnipers.com/top-10-largest-mobile-companies-in-the-world-2018/}}. Table~\ref{tab:top-10_companies} presents the list of top-10 companies in each of the above three segments.   

\begin{table}[!tbh]
      \centering
        \begin{tabular}{c|c|c}
          \textbf{Software} & \textbf{Hardware} & \textbf{Mobile Phone}\\
          \hline
          Microsoft & Apple & Samsung \\
          Google & Samsung & Apple \\
          IBM & IBM & Huawei \\
          Oracle & Foxconn & Oppo \\
          Facebook & Hewlett Packard & Vivo \\
          Tencent & Lenovo & Xiaomi \\
          SAP & Fujitsu & OnePlus \\
          Accenture & Quanta Computer & Lenovo \\
          TCS & AsusTek & Nokia \\
          Baidu & Compal & LG \\
        \end{tabular}
        \caption{ \label{tab:top-10_companies}Top-10 \textit{Software}, \textit{Hardware}, and \textit{Mobile Phone} companies selected from three publicly available lists.}
\end{table}

In subsequent sections, we analyze patents filed by these companies over the years.  In our patent dataset, each company can have several variations in name due to multiple research groups, geographical locations, subsidiaries, headquarters, etc. For example, IBM is present as `International Business Machines Corporation Armonk', `International Business Machines Laboratory Inc.', etc. We overcome these inconsistencies by manually annotating name variations. However, we claim that basic string matching techniques can easily automate this normalization. Besides, we eliminate frequently occurring catchphrases like, 'method', 'present invention', etc., to ignore noisy/redundant signals. This filtering process was automated by removing catchphrases with top-10 document frequencies. We next, present how catchphrases can be leveraged in understanding the topical evolution of companies. 

\subsection{Topic evolution}
\label{sec:jaccard}
\begin{figure*}[!tbh]
    \includegraphics[width=\hsize, keepaspectratio]{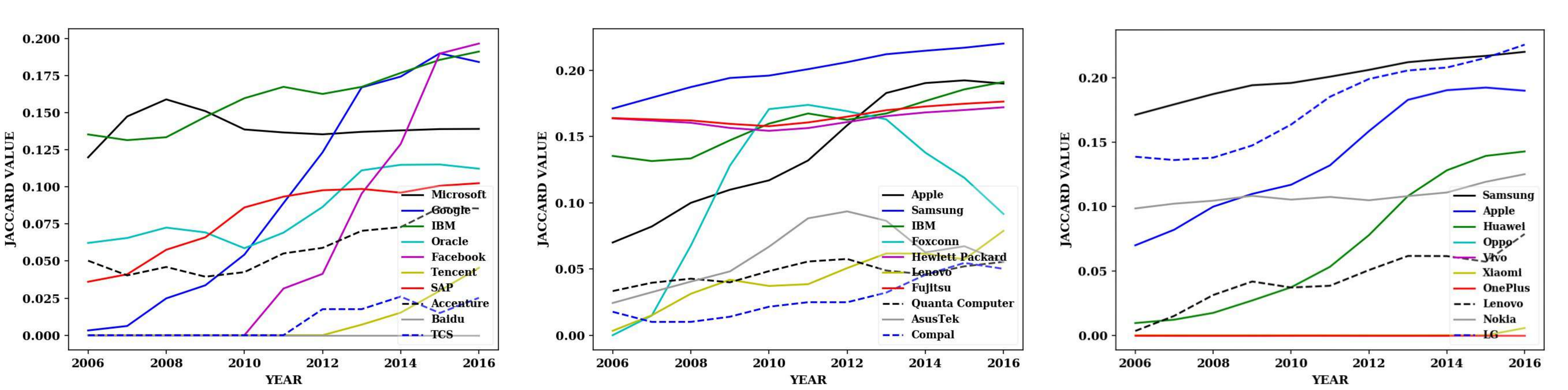}
    \caption{Moving average of catchphrases similarity between consecutive years for -- \textit{Software} (left), \textit{Hardware} (center), and \textit{Mobile Phone} (right) companies.}\label{fig:jaccard_plot}
\end{figure*}

In this section, we study the topical evolution of companies. We leverage the \textit{Jaccard Similarity} (JS) between the catchphrases to compute the topical overlap between patents filed in consecutive years by a specific company. We conduct this experiment for 11 years between 2006--2016. Figure~\ref{fig:jaccard_plot} shows temporal profiles of a three-year moving average over JS for each of the three segments. We observe that Baidu in \textit{Hardware} segment while Oppo, Vivo, and OnePlus in \textit{Mobile Phones} segments exhibit relatively low similarity between catchphrases over the years. However, most of the companies have similarity curves with multiple peaks with an overall increase in the JS values over the years. For this analysis, we only considered 2-gram catchphrases. However, we found similar observations for higher n-gram catchphrases. If an organization is filing patents on the same topics over the years, the JS value will only increase; on the other hand, if an organization is continuously filing patents on newer topics, the JS value is expected to decline.

\subsection{Categorization}
Further, we conduct a nuanced study to understand this temporal behavior. We classify each company's similarity profile into five categories~\cite{chakraborty2015categorization} based on the number and location of peaks. A peak in the similarity profile of a company represents a high topical similarity between consecutive years followed by a topical drifting off period. We leverage the peak identification method proposed by~\citet{chakraborty2015categorization}. Note that peaks occurring in consecutive years are considered as a single peak. The categories are:
\begin{enumerate}
    \item \textsc{MonInc}: Similarity profile that monotonically increases. The peak occurs in the last year.
    \item \textsc{MonDec}: Similarity profile that monotonically decreases. The peak occurs in the first year.
    \item \textsc{PeakInit}: Similarity profile that consists single peak within the first three years but not the first year.
    \item \textsc{PeakLate}: Similarity profile that consists single peak after the initial three years but not the last year.
    \item \textsc{PeakMult}: Similarity profile consisting of multiple peaks. 
    \item \textsc{Others}: Similarity profiles that do not qualify into the above categories are kept in this category. They mainly consist of profiles with extremely low JS values for each year.
\end{enumerate}

Table~\ref{tab:category} shows categorization results. We find no company in \textsc{MonDec} and \textsc{PeakInit} categories. Majority of the companies are present in the \textsc{PeakMult} category followed by \textsc{PeakLate} category. Companies in \textsc{Others} category have very less number of filed patents. Three out of four companies in \textsc{Others} category are recently launched mobile companies.

\begin{table}[!tbh]
    \begin{center}
    \begin{tabular}{l|c|l}
      \textbf{Category} & \textbf{Count} & \textbf{Names}\\
      \hline
      \textsc{MonInc} & 4 & Tencent, Samsung, Xiaomi, Lenovo \\
      \textsc{MonDec} & 0 &  \\
      \textsc{PeakInit} & 0 &  \\
      \textsc{PeakLate} & 6 & \parbox[t]{5.5cm}{Facebook, TCS, Huawei, AsusTek, Foxconn, Compal }\\
      \textsc{PeakMult} & 12 & \parbox[t]{5.5cm}{HP, SAP, Accenture, Nokia, Fujitsu, Quanta Computer, Microsoft, IBM, Oracle, Google, Apple, LG} \\
      \textsc{Others} & 4 & Baidu, Oppo, Vivo, OnePlus \\
    \end{tabular}
    \caption{\label{tab:category}Categorization of top-10 \textit{Software}, \textit{Hardware}, and \textit{Mobile Phone} companies based on temporal catchphrase similarity profile. No company was classified in \textsc{MonDec} and \textsc{PeakInit} category.}
    \end{center}
\end{table}

Even though, \textsc{PeakMult} category consists multiple peaks, we observe two distinct fluctuation patterns. We term these patterns as (i) \textsc{stable} and (ii) \textsc{unstable}. In \textsc{stable}, the profile looks considerably less fluctuating. The profile highly fluctuates in \textsc{unstable} category. We quantify the above fluctuating patterns by leveraging the average value of JS. Given, JS(c) is the similarity profile for a company $c$, average value of JS ($avg_{JS} (c)$) is computed as:

\begin{equation}
avg_{JS}(c) = \frac {min(JS(c)) + max(JS(c))} {2}
\end{equation}

Empirically, we observe that companies with $avg_{JS} > 0.1$ can be classified as \textsc{unstable}, while the rest can be classified as \textsc{stable}. Table~\ref{tab:peakmult} shows companies in the \textsc{PeakMult} category that are further categorized into \textsc{stable} and \textsc{unstable}. Among, \textsc{stable} and \textsc{unstable} sub-categories, the former contains more (=7) companies than the latter (=5). 

\begin{table}[!tbh]
      \begin{center}
        \begin{tabular}{c|c|c}
          \textbf{Company} & \textbf{$avg_{JS}$} & \textbf{Category}\\
          \hline
            Nokia & 0.040 & \textsc{stable} \\
            Fujitsu & 0.085 & \textsc{stable} \\
            Quanta Computer & 0.069 & \textsc{stable} \\
            Microsoft & 0.105 & \textsc{stable} \\
            Accenture & 0.040 & \textsc{stable} \\
            SAP & 0.048 &\textsc{stable} \\
            Hewlett Packard & 0.084 & \textsc{stable} \\ 
            LG & 0.223 & \textsc{unstable} \\
            Oracle & 0.117 & \textsc{unstable} \\
            Google & 0.121 & \textsc{unstable} \\
            Apple & 0.197 & \textsc{unstable} \\
            IBM & 0.124 & \textsc{unstable}\\
        \end{tabular}
        \caption{\label{tab:peakmult}List of companies in \textsc{PeakMult} that are classified into \textsc{stable} and \textsc{unstable} sub-categories along with the average value of Jaccard  Similarity ($avg_{JS}$) used for categorization.}
      \end{center}
\end{table}

\subsection{Citation count}
Citations, in the scholarly world, determine the popularity of research papers/authors/organizations. Here, we adopt a similar analogy for patent articles.  A patent citation is a document cited by an applicant, third party, or a patent office examiner because its content relates to a patent application. We compute the citation count of a patent $p$ by summing the citations received by $p$. For the current study, we construct citer-cited pairs by extracting references present in patent texts and use these pairs to compute patent citation counts. 

Next, we create multiple citation zones based on the citation count of a patent. We define four distinctive zones: (i) very low, (ii) low, (iii) medium, and (iv) high, to study the influence of the JS profile of a company on the number of citations received by its patents.  Table~\ref{tab:zoning}  presents zoning statistics of the complete dataset. Out of 3,829,153 patent articles, 1,499,175 have zero citation count. 

\begin{table}[!tbh]
  \begin{center}
    \begin{tabular}{c|c|c}
      \textbf{Category} & \textbf{Citation Count} & \textbf{Patent Count}\\
      \hline
      Very Low &0&1,499,175\\
      Low & 0$<$ x $<$5 & 1,274,029 \\
      Medium & $5\leq$ x $<$25 & 840,461 \\
      High & x$\geq$25  & 215,488 \\
    \end{tabular}
    \caption{\label{tab:zoning}Patent citation zones with distinct citation count ranges.}
  \end{center}
\end{table}

\begin{figure}[!tbh]
    \begin{tabular}{@{}c@{}c@{}}
        \includegraphics[width=\hsize]{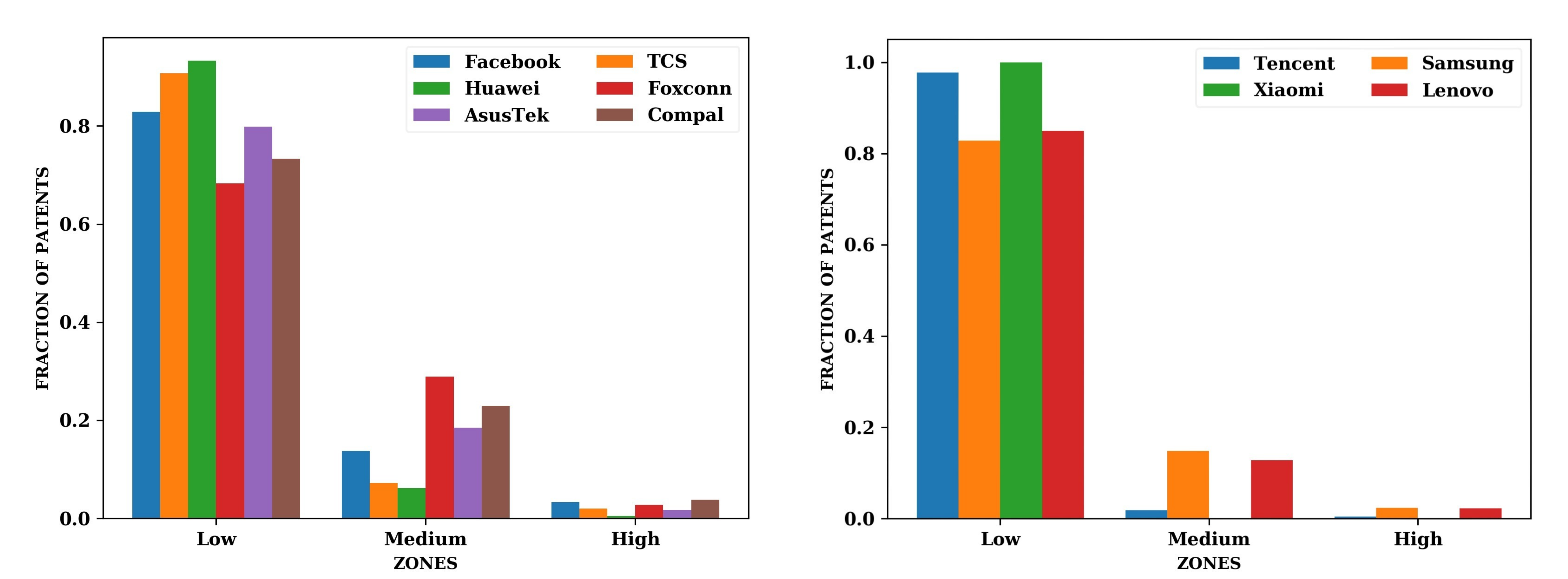}
    \end{tabular}
    \caption{\label{fig:peaklate_moninc}Citation count zones vs similarity profiles: Fraction of patents in \textsc{PeakLate} (left) and \textsc{MonInc} (right) category companies in each citation count zone.}
\end{figure}

Next, we relate similarity profiles and citation count zones. For each company, we measure the fraction of patents in different citation zones. We leverage histograms as a visualization tool to conduct this study. In Figure~\ref{fig:peaklate_moninc}, we observe that the fraction of patents in \textit{Medium} and \textit{High} citation zones in \textsc{PeakLate} category are relatively higher than in \textsc{MonInc} category. This indicates that the introduction of diversity in topics over time helps in enhancing the future citations of the patents filed by a company.

\begin{figure}[!tbh]
    \begin{tabular}{@{}c@{}c@{}}
        \includegraphics[width=\hsize]{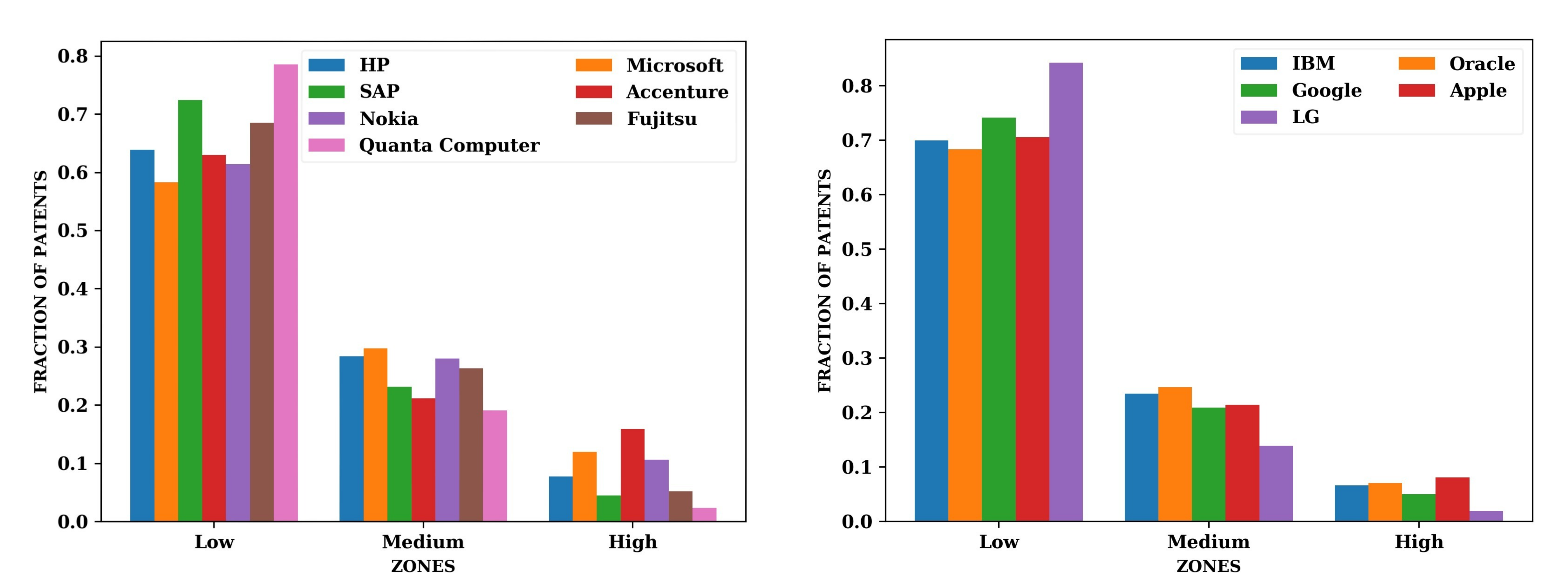}
    \end{tabular}
    \caption{\label{fig:stable_unstable}Citation count zones vs similarity profiles: Fraction of patents in \textsc{stable} (left) and \textsc{unstable} (right) category companies in each citation count zone.}
\end{figure}

Figure~\ref{fig:stable_unstable} compares two subcategories of \textsc{PeakMult}. We observe that the fraction of patent falling under the \textit{Medium}, and \textit{High} citation zones in \textsc{unstable} category is relatively higher than \textsc{stable} categories implying that the companies with high fluctuations in similarity profiles perform better in terms of receiving citation counts. A possible explanation is that the companies with relatively specialized research domain file patents which attract lesser citations than the companies with diversified research domain.

\begin{figure}[!tbh]
\begin{tabular}{@{}c@{}c@{}}
    \includegraphics[width=0.8\hsize]{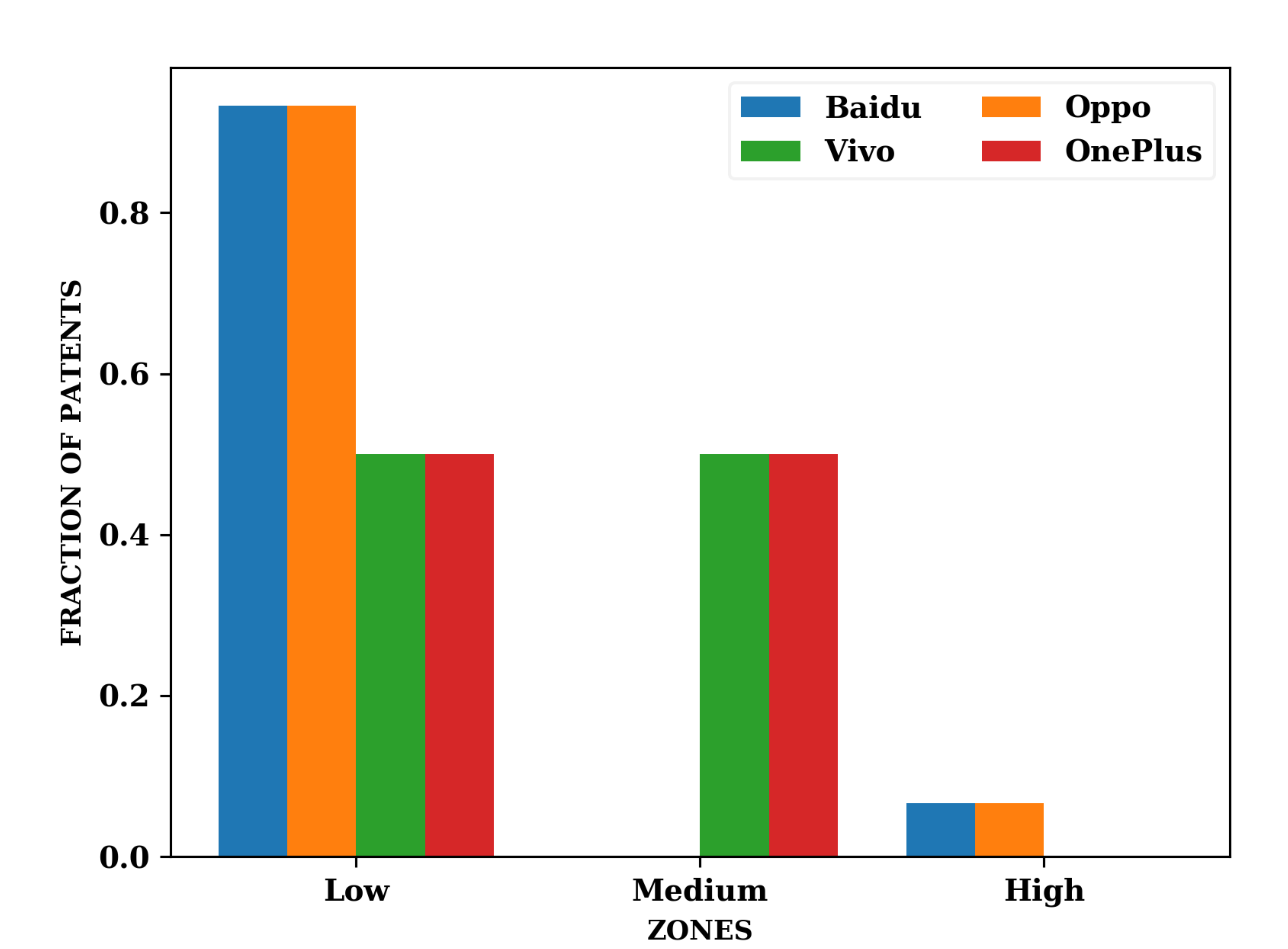}
\end{tabular}
\caption{Citation count zones vs similarity profiles: Fraction of patents in \textsc{Others} in each citation count zone.}\label{fig:others}
\end{figure}

Lastly, we study \textsc{Others} category in Figure~\ref{fig:others}. Quite surprisingly, we observe that the fraction of patents in \textit{Medium} and \textit{High} citation zones in \textsc{Others} category is relatively higher than the rest of the categories described above in Figures~\ref{fig:peaklate_moninc} and \ref{fig:stable_unstable}.

\begin{figure*}[!tbh]
    \centering
    \includegraphics[width=0.78\hsize, keepaspectratio]{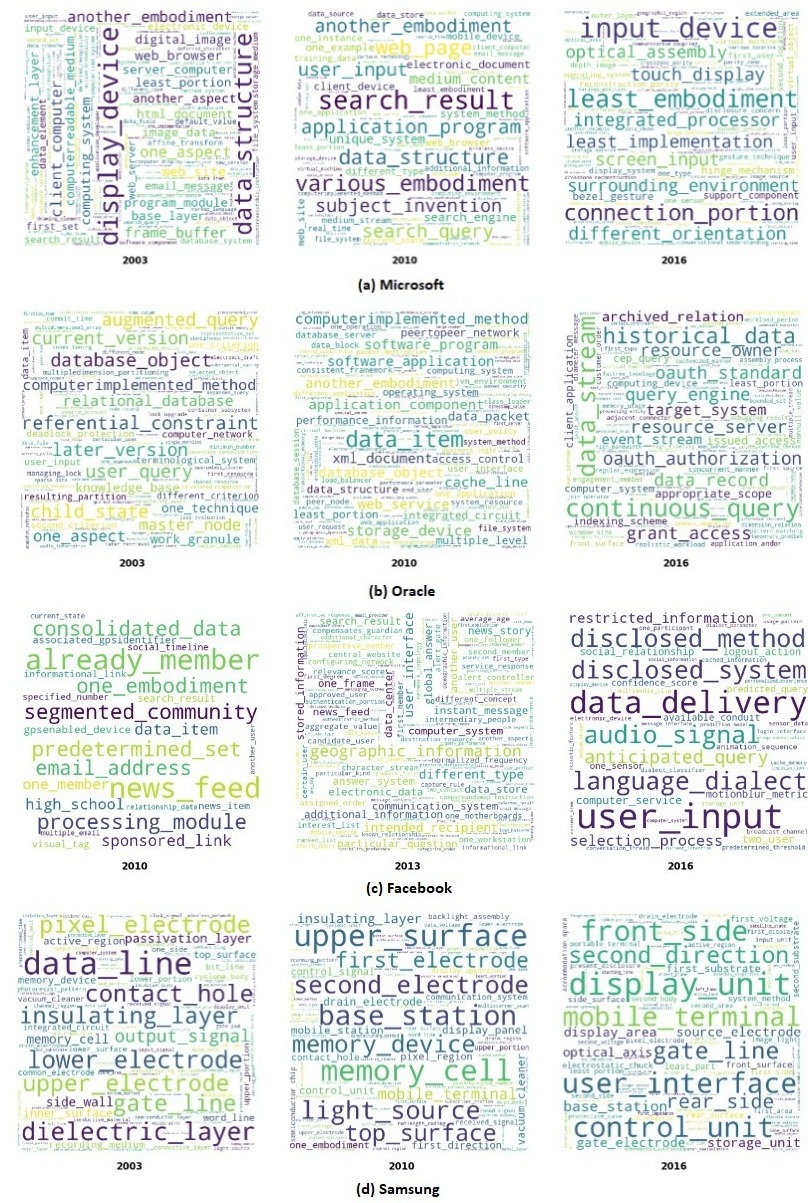}
        \caption{Word clouds of the representative companies in different similarity profile based categories at three distinct years. (a)~\textsc{stable} (Microsoft), (b)~\textsc{unstable} (Oracle), (c)~\textsc{peaklate} (Facebook), and (d)~\textsc{moninc} (Samsung). }\label{fig:word_cloud}
\end{figure*}

\subsection{Catchphrases in the \textsc{stable} and \textsc{unstable} groups}
In this section, we analyze the extent of usage of certain catchphrases (bigrams and trigrams) by a company. We rank the catchphrases based on document frequency, i.e, the number of patent documents a catchphrase is present in. Tables~\ref{tab:stablebigrams} and ~\ref{tab:unstablebigrams} show the top-10 bigrams for companies present in the \textsc{stable} and \textsc{unstable} groups respectively. Table~\ref{tab:stabletrigrams} and ~\ref{tab:unstabletrigrams} show top-10 trigrams for the same companies. Last, Table~\ref{tab:stable-unstable} notes the top-10 bigrams and trigrams from the entire stable and unstable categories taken together. While the \textsc{stable} group is concerned more about computer systems, the \textsc{unstable} group is more about electronic device parts.

\begin{table*}[t]
    \resizebox{\textwidth}{!}{
    \begin{tabular}{*{6}{c|} c}
     \textbf{HP} &\textbf{MS} &\textbf{SAP} &\textbf{Accenture} &\textbf{Nokia} &\textbf{Fujitsu} &\textbf{Quanta Computer} \\
     \hline
        print job & client device & business process & processing device & user interface & closed position & circuit board \\
         one aspect & search result & application server & third party & communication device & inner surface & display panel \\
         printing system & application program & application program & real-world environment & one embodiment & opposite side & second image \\
         second set & user input & software application & mobile device & computer program & upper surface & one side \\
         operating system & computing system & business application & invention concern & telecommunication system & longitudinal axis & second end \\
         second side & search engine & system method & educational material & telecommunication network & another embodiment & battery module \\
         second position & data store & data structure & computer-implemented method & access point & opposite end & second position \\
         second portion & least portion & system software & communication network & data transmission & open position & one end \\
         display device & subject matter & user input &synchronized video & least part & bottom surface & portable computer \\
         present disclosure & client computer & business object & solution information & second device & side wall & power supply \\
    \end{tabular}
    }
    \caption{\label{tab:stablebigrams}Bi-grams with the top 10 document frequency values in STABLE category.}
\end{table*}

\begin{table*}[t]
    \centering
    \resizebox{\textwidth}{!}{
    \begin{tabular}{*{4}{c|} c}
     \textbf{IBM} &\textbf{Oracle} &\textbf{Google} &\textbf{Apple} &\textbf{LG} \\
     \hline
        top surface & application server & user interface & integrated circuit & second electrode \\
         operating system & operating system & present disclosure & second set & one side \\
         storage device & data structure & one example & user input & common electrode \\
         computer program & second set & system method & one example & light source \\
         second set & software application & example method & first set & lcd device \\
         computing system & one technique &content item & operating system & control information \\
         another embodiment & source code & user input & another embodiment & display device \\
         user interface & computer-implemented method & one processor & least portion & array substrate \\
         drain region & another aspect & user device & host device & drain electrode \\
         data structure & database object & subject matter & client device & washing machine \\
    \end{tabular}
    }
    \caption{\label{tab:unstablebigrams}Bi-grams with the top 10 document frequency values in UNSTABLE category.}
\end{table*}

\begin{table*}[t]
    \centering
    \resizebox{\textwidth}{!}{
    \begin{tabular}{*{6}{c|} c}
     \textbf{HP} &\textbf{MS} &\textbf{SAP} &\textbf{Accenture} &\textbf{Nokia} &\textbf{Fujitsu} &\textbf{Quanta Computer} \\
     \hline
        storage area network & host operating system &first data object &dual information system &first base station & user 's head & portable electronic apparatus \\
         least one component &client computing device & one general aspect & telecommunication industry taxonomy &packet data network &first second portion & mobile communication device \\
        fluid ejection assembly &mobile communication device & business process model & contact center representative & least one parameter & user 's foot & second frequency band \\
        first second set & user 's interaction & second user input & contact center system &first network element & least one opening & third conductor arm \\
        least one component & least one implementation &least one service & context-appropriate enforcing completion & user equipment due & thinning spraying irrigation & portable computer system \\
        disclosed embodiment relate & distributed computing system & core software platform &location-based service system & wireless communication device & patient 's body &second radiating element \\
         least one surface & application program interface & least one attribute & cognitive educational experience & least one cell & least one side & blade server system \\
         inkjet ink composition &one computing device & second data object &individualized learning experience &wireless communication system & least one aperture &service agent server \\
        central processing unit & client computer system &one exemplary embodiment & user 's comprehension &wireless communication device &storied index rating &printed circuit board \\
         graphical user interface & wireless access point & related method system & object recognition analysis & second base station & usda hardiness zone & wireless communication device \\
    \end{tabular}
    }
    \caption{\label{tab:stabletrigrams}Tri-grams with the top 10 document frequency values in STABLE category.}
\end{table*}

\begin{table*}[t]
    \centering
    \resizebox{\textwidth}{!}{
    \begin{tabular}{*{4}{c|} c}
     \textbf{IBM} &\textbf{Oracle} &\textbf{Google} &\textbf{Apple} &\textbf{LG} \\
     \hline
         first conductivity type &current result list & one search result & electronic device housing &light guide plate \\
         field effect transistor & flexible extensible architecture & first search result & scrolling 3d manipulation & digital broadcasting system \\
         second dielectric layer &computer program product & image sensor interface & intuitive hand configuration & second semiconductor layer \\
         gate dielectric layer & distributed computing environment & disclosed subject matter & hand approach touch & liquid crystal cell \\
         data communication network & graphical user interface & image search result &proximity-sensing multi-touch surface &light emitting diode \\
         integrated circuit device & data storage system &client computing device & wireless communication circuitry & main service data \\
         direct physical contact &data processing system & client computing device &antenna resonating element & image display device \\
         second conductivity type & application programming interface &second computing device & computer readable medium & serving base station \\
         database management system & database management system &mobile communication device & wireless communication system & light emitting diode \\
         buried insulator layer & contention management mechanism & distributed storage system & wireless electronic device &first second electrode \\
    \end{tabular}
    }
    \caption{\label{tab:unstabletrigrams}Tri-grams with the top 10 document frequency values in UNSTABLE category.}
\end{table*}

\begin{table*}[t]
    \centering
    \begin{tabular}{c|c||c|c }
     \multicolumn{2}{c||}{\textbf{STABLE}} & \multicolumn{2}{c}{\textbf{UNSTABLE}} \\
     \textbf{Bi-grams} &\textbf{Tri-grams} &\textbf{Bi-grams} &\textbf{Tri-grams} \\
     \hline
          closed position &first second portion & top surface & second semiconductor layer \\
         another embodiment & user's head & user interface & printed circuit board \\
         opposite side & least one opening & second set &light guide plate \\
         inner surface & least one side & operating system &light emitting diode \\
         upper surface & user's foot & present disclosure & digital broadcasting system \\
         one aspect & least one aperture & least portion &first semiconductor layer \\
         longitudinal axis & patient's body & another embodiment & light emitting diode \\
         open position & thinning spraying irrigation & system method & liquid crystal cell \\
         second position & central processing unit & data structure &first second electrode \\
         opposite end &storie index rating & computing system & serving base station \\
    \end{tabular}
    \caption{\label{tab:stable-unstable}Bi-grams and Tri-grams with the top 10 document frequency values.}
    \vspace{-0.5cm}
\end{table*}

\subsection{Temporal visualizations} 
In this section, we study the catchphrase evolution of companies. As a popular visualization tool, we leverage \textit{word clouds}. We create word clouds for each company between the years 2003--2016. Due to space constraints, in Figure~\ref{fig:word_cloud}, we only consider word clouds for one representative company from \textsc{stable}, \textsc{unstable}, \textsc{peaklate} and \textsc{moninc} categories at three representative years. We claim that catchphrase evolution presents a fair understanding of the changing innovation trends of companies. Note that we consider only bigram catchphrases in this study. We can conduct a similar study for any company in different years\footnote{The detailed word clouds for all companies in our dataset are available at \url{http://tinyurl.com/y5ynhj9n}}. 

In Figure~\ref{fig:word_cloud}a, we study catchphrase evolution for Microsoft (a representative company in the \textsc{stable} group). We observe a shift from traditional topics such as client-server models, databases, basic Web development, etc. (in 2003), toward full-fledged Web search and Internet technologies (in 2010). In 2016, the focus shifted to mobile devices and gesture identification. The above trends coincide with several product releases such as BING (a search engine released in 2009)\footnote{https://en.wikipedia.org/wiki/Bing\_(search\_engine)} and Lumia (mobile phones released in 2015)\footnote{https://en.wikipedia.org/wiki/Microsoft\_Lumia\_435}.

In Figure~\ref{fig:word_cloud}b, we study catchphrase evolution for Oracle (a representative company in the \textsc{unstable} category). Oracle seems to have shifted its focus from traditional database topics like relational databases, query, etc. (in 2003), toward the development of software as a service (SAAS) in 2010. In 2016, it continued to focus on services with a major emphasis on reliable authentication mechanisms in the cloud. These innovation trends resulted in several products like \textit{Oracle cloud} (cloud computing service launched in 2016), Primavera (an enterprise project portfolio management software acquired by Oracle in 2008), etc. 

Similarly, in Figure~\ref{fig:word_cloud}c, we study catchphrase evolution for Facebook (a representative company in \textsc{PeakLate} category). As Facebook started its operations from 2004, we present visualizations for three years, 2010, 2013 and 2016. The initial focus was to develop technical features like news feeds, membership, etc. In the year 2013, these trends shift toward instant messaging aspects. In the year 2016, the catchphrases show a distinct innovation pattern of restricting and disclosing data availability. Facebook Messenger (introduced in 2011) is one of the products developed between 2011--2013\footnote{https://en.wikipedia.org/wiki/Facebook}.

We study Samsung as a representative company in \textsc{MonInc} category (see Figure~\ref{fig:word_cloud}d). Primarily Samsung's major focus lies in traditional electronics innovation. Recent trends suggest an increased focus on mobile technologies such as user interfaces, display units, etc. 

\section{Conclusion and future work}
\label{sec:con}
In this paper, we propose an unsupervised catchphrase identification and ranking system. Our proposed system achieves a substantial improvement, both in terms of
precision and recall, against state-of-the-art techniques. We demonstrate the usability of this extraction by analyzing how topics evolve in patent documents and how these evolution patterns shape the future citation count of the patents filed by a company.

In the future, we plan to extend the current work by developing an online interface for automatic catchphrase identification. We also plan to understand the influence of catchphrase evolution on the company's revenue.

\bibliographystyle{ACM-Reference-Format}
\bibliography{ref}
\end{document}